\documentclass[a4paper]{cas-dc}

\usepackage[numbers]{natbib}
\usepackage{nicefrac, xfrac}

\usepackage{amsmath}

\usepackage{xcolor}
\usepackage{soul}

\def\tsc#1{\csdef{#1}{\textsc{\lowercase{#1}}\xspace}}
\tsc{WGM}
\tsc{QE}
\tsc{EP}
\tsc{PMS}
\tsc{BEC}
\tsc{DE}

\begin{document}
\let\WriteBookmarks\relax
\def\floatpagepagefraction{1}
\def\textpagefraction{.001}

\shorttitle{}    

\shortauthors{E. Vlasov et~al.}

\title [mode = title]{Secondary electron topographical contrast formation in scanning transmission electron microscopy}

\author[1]{Evgenii Vlasov}

\credit{Writing - original draft, Conceptualization, Methodology, Software}

\affiliation[1]{organization={EMAT and NANOlight Center of Excellence, University of Antwerp},
            addressline={Groenenborgerlaan 171}, 
            city={Antwerp},
            postcode={2020},
            country={Belgium}}

\author[1]{Wouter Heyvaert}
\credit{Conceptualization, Methodology, Software}

\author[1]{Tom Stoops}
\credit{Conceptualization, Methodology}

\author[1]{Sandra Van Aert}
\credit{Writing - review \& editing, Supervision}

\author[1]{Johan Verbeeck}
\credit{Writing - review \& editing, Supervision, Conceptualization}

\author[1]{Sara Bals}
\credit{Writing - review \& editing, Supervision, Conceptualization}
\cormark[1]
\ead{sara.bals@uantwerpen.be}

\cortext[cor1]{Corresponding author}

\begin{abstract}
Secondary electron (SE) imaging offers a powerful complementary capabilities to convetional scanning transmission electron microscopy (STEM) by providing surface-sensitive, pseudo-3D topographic information. However, contrast interpretation of such images remains empirical due to complex interactions of emitted SE with the magnetic field in the objective field of TEM. Here, we propose an analytical physical model that takes into account the physics of SE emissin and interaction of the emitted SEs with magnetic field. This enables more reliable image interpretation and potentially lay the foundation for novel 3D surface reconstruction algorithms.
\end{abstract}

\begin{highlights}
\item Developed an analytical physical model of topographical SE contrast formation in STEM.
\item SE-STEM topographical contrast artifacts can be analyticaly predicted.
\item SE-STEM contrast is a function of the surface inclination angle and local thickness of the specimen.
\end{highlights}


\maketitle

\section{Introduction}

Scanning Transmission Electron Microscopy (STEM) is a preferred imaging technique in the field of material science~\cite{vantendeloo2012, guzzinati2018}. STEM images are formed from elastically scattered electrons, transmitted through a thin specimen, providing both crystallographic (low-angle scattering) and mass-thickness (high-angle scattering) information~\cite{williams2009}. Conventionally, STEM images represent only a two-dimensional (2D) projection of the sample under investigation, and therefore, convey limited direct information about the surface morphology, which may hamper an accurate understanding of the three-dimensional (3D) structure of the material. Whereas the 3D structure may sometimes be inferred from image contrast in the case of regularly shaped nanoparticles (NPs), it becomes significantly more challenging for NPs with irregular shapes. This limitation is particularly important, as the intricate 3D structures of NPs are closely related to their properties, which are relevant in \textit{e.g.} plasmonic~\cite{seplveda2009, atwater2010} or catalytic applications~\cite{li2010}. \par

Secondary electron STEM (SE-STEM) imaging is considered a powerful complementary technique to conventional STEM imaging that yields pseudo-3D surface (topographic) information~\cite{mitchell2016,sangabriel2024}. The SEs are generated during inelastic scattering events, when the incident electrons transfer their energy to the atomic electrons in the sample, leading to the emission of SEs. A low energy of SEs ($\leq$~50~eV) limits the depth from which they can escape from the specimen to a few nanometers from the surface making the SE signal highly surface-sensitive~\cite{reimer1998}. SE imaging has been incorporated into STEM for several decades~\cite{allen1982,hembree1989,liu1991,howie1995,zhu2009,inada2014}. The higher acceleration voltages, compared to scanning electron microscope (SEM), the use of ultra-thin samples and introduction of abberation-correctors enabled SE imaging with resolution down to the sub-nm or even atomic resolution regime~\cite{zhu2009,inada2014,ciston2015}. It was demonstrated that SE-STEM can complement conventional STEM imaging modes, offering access to the depth and topography information inaccessible by conventional STEM imaging~\cite{mitchell2016}. It greatly aids interpretation of the complex contrast in STEM images that can arise in topographically and compositionally diverse components. For example, SE-STEM enables direct observation of NP surfaces, making it particularly attractive for analyzing the morphologies of complex NPs with significantly improved throughput compared to electron tomography.  Recently we have shown that SE-STEM imaging can be used to quantify the helical morphology of chiral Au nanorods with high-throughput, enabling the correlation between ensemble-level morphological characteristics with chiroptical properties~\cite{vlasov2024}. Additionally, the SE-STEM imaging is interesting when studying nanomaterials in realistic conditions. SE-STEM has been used in combination with environmental TEMs to study the dynamic behaviour and degradation mechanisms of supported cataysts under reaction conditions (gas atmosphere and high temperatures)~\cite{sangabriel2024}.\par

Given the limited availability of dedicated SE detectors in modern STEM instruments, recently, an alternative approach for SE-STEM imaging, using electron induced beam current (SEEBIC), has been proposed~\cite{hubbard2018}. The SEEBIC technique is based on the measurement of an electrical current induced by the emission of SEs from an electron-transparent sample. The SEEBIC signal is proportional to the number of emitted SEs and therefore enables direct measurements of SE yield~\cite{feng2024}. SEEBIC can be considered as an attractive approach to image the morphology of nanomaterials with shorter acquisition and processing times in comparison to electron tomography (a few minutes \textit{vs.} up to an hour) and superior resolution compared to SEM~\cite{vlasov2023}.\par

Despite demonstrated capabilities to directly reveal surface morphology at high spatial resolution, the interpretation of SE-STEM contrast remains largely empirical and is not always straightforward due to interaction of emitted SEs with the magnetic field of the STEM objective lens~\cite{vlasov2023}. In order to make SE-STEM imaging a reliable tool for studying the morphology of nanomaterals, a thorough understanding of the mechanisms of topographical SE contrast formation in STEM is required. Although SE images provide only a pseudo-3D perception, we foresee that the third dimension could be retrieved from single images given rapid development of modern machine learning algorithms~\cite{nerf,dreamgaussian,pix2mesh}. However, such 3D reconstructions would require accurate forward models of SE-STEM contrast formation. In this paper, we propose an analytical model of topographical contrast considering the interaction of emitted SEs with the magnetic field in the objective lens of STEM.

\section{Physics of SE imaging in STEM}

SE emission is a complex process that is a result of a sequence of physical events~\cite{reimer1998,egerton2022}. First, primary electrons scatter inelastically and transfer their energy to electrons of the sample of interest. This energy transfer creates SEs by excitation of core and conduction electrons and decay of plasmons. Produced SEs diffuse through the solid with a kinetic energy ranging from a few eV to hundreds of eV and slow down until they are absorbed or reach an external surface. As a result, only those electrons that are produced near the surface can escape into the vacuum. However, the SEs can only be detected by an external detector if their kinetic energy allows them to surmount the surface-potential barrier, whose height is related to the work function of the specimen. This entire procees ($S$) can be described as~\cite{inada2014,egerton2022}:
\par

\begin{equation} \label{eq:Equation 1}
    S = G \cdot T \cdot B \cdot D,
\end{equation} \par

\noindent{}where term $G$ represents the SE generation in the sample, $T$ describes the diffusion (transmission) of those SEs to the surface, of which a fraction $B$ overcome the surface barrier and are detected by a detector with an efficiency $D$. \par

The intensity change in the SE image scanned along a line or, in other words, SE image contrast then is defined by:

\begin{equation} \label{eq:Equation 2}
    \frac{dS}{dx} = 
    T B D \frac{dG}{dx} + G B D \frac{dT}{dx} + G T D \frac{dB}{dx} + G T B \frac{dD}{dx}.
\end{equation} \par

Here, for simplicity, factors $G$, $T$ and $B$ are treated as independent functions, although in practice they are all related to the underlying electronic structure of the material. The first term provides an atomic-number or stopping power contrast in atomic-resolution SE-STEM images (if specimen is chemically inhomogeneous~\cite{inada2014}), while the second one serves as a source of topographic contrast. The third term conveys the work function contrast, and the fourth term can provide the voltage SE contrast. In this paper we focus primarily on topographic contrast in SE-STEM imaging.

\section{Physical model of topographic contrast}

Most of the generated SEs have energies below 100~eV, whereas both elastic and inelastic mean free paths are of the order of a few nanometers, so they are strongly scattered within the material. SEs rapidly slow down during inelastic scattering events until they are absorbed as a conduction or valence electrons. As a result, the SEs can escape the specimen only form a shallow escape depth $t_{SE}$, which is in the range of 0.5-1.5~nm for metals and increases to in the range of 10-20~nm for insulators~\cite{reimer1998}. The probability of escape form depth $z$ below the surface can be described as follows~\cite{reimer1998}:

\begin{equation} \label{eq:Equation 3}
    P \left( z \right) \approx P \left( 0 \right) \exp{\left( -\frac{z}{t_{SE}} \right)}.
\end{equation} \par

From experimental observations in SEM, it is known that the SE yield $\delta$ increases with increasing specimen tilt angle $\alpha \in [0, \frac{\pi}{2}]$ as $\sec{(\alpha)}$. This is attributed to an increase of the path length of the primary electron inside the exit depth, which is equal to $t_{SE} \sec{(\alpha)}$. On the other hand, the angular distribution $\frac{d \delta}{d \Omega}$ of the emitted SEs is experimentally observed to be proportional to $\cos{(\theta)}$, where $\theta$ denotes the emission angle relative to the surface normal, known as Lambert's cosine law. The dependence of SE yield $\delta$ on $\alpha$ and $\theta$ can, therefore, be described as~\cite{reimer1998}:

\begin{equation} \label{eq:Equation 4}
    \frac{d \delta}{d \Omega} = \frac{\delta_{0}}{\pi} \sec{(\alpha)} \cos{(\theta)},
\end{equation} \par 

\noindent{}where material-dependent coefficent $\delta_{0} = \delta_{0}(Z)$ denotes the SE yield for normal incidence ($\alpha = 0$) for a given material.

Integration of $\frac{d \delta}{d \Omega}$ over a solid angle $d \Omega$ gives the total SE yield (assuming typical SEM conditions: SE emission from the top surface of the sample and the location of the SE detector above the sample):

\begin{equation} \label{eq:Equation 5}
    \delta = \frac{\delta_{0}}{\pi} \sec{(\alpha)} 
    \int_{0}^{2 \pi} d \varphi \int_{0}^{\frac{\pi}{2}} d \theta\, \sin{(\theta)} \cos{(\theta)},
\end{equation} \par

\noindent{} resulting in

\begin{equation} \label{eq:Equation 6}
    \delta = \delta_{0} \sec{(\alpha)}, \alpha \in \left[0, \frac{\pi}{2}\right).
\end{equation} \par

\textbf{Equation~\ref{eq:Equation 6}} well describes SE emission in SEM and is often used for interpretation and simulation of SEM images. Moreover, it can even be used for shape-from-shading 3D reconstructions~\cite{zhu2014}. However, it is poorly applicable in the case of SE-STEM (or SEEBIC) imaging due to several factors. First, in STEM conditions (e-beam energies $\geq$ 60~keV and ultra-thin samples), the primary electron beam traverses the samples, leading to the presence of both entrance and exit surfaces in the specimen that emit SEs. Second, the presence of a strong magnetic field inside the TEM pole-piece gap strongly influences the trajectories of emitted SEs. These effects can strongly impact SE image formation and therefore require thorough consideration.

\section{Problem statement}

To address the topographic SE contrast formation in STEM, we consider the following problem: a NP is placed on a continuous support that is conductive (in case of SEEBIC imaging) and absorbing (meaning that the thickness of the support exceeds an elastic and inelastic mean free paths for the SEs) in a homogeneous strong magnetic field ($\approx$ 2~T) in the TEM pole piece gap. For the sake of simplicity, we, here, only consider convex-shaped NPs, meaning that all internal angles are less than 180$^{\circ}$ and there is no recapture of SEs by sharp tips and protrusions present in the NP itself. The evaluation of the recapture of SEs by the structural features of the NP would require finding a solution for a ray tracing problem, which is out of scope of the current paper. \par

\begin{figure}
	\centering
	\includegraphics[width=1\columnwidth]{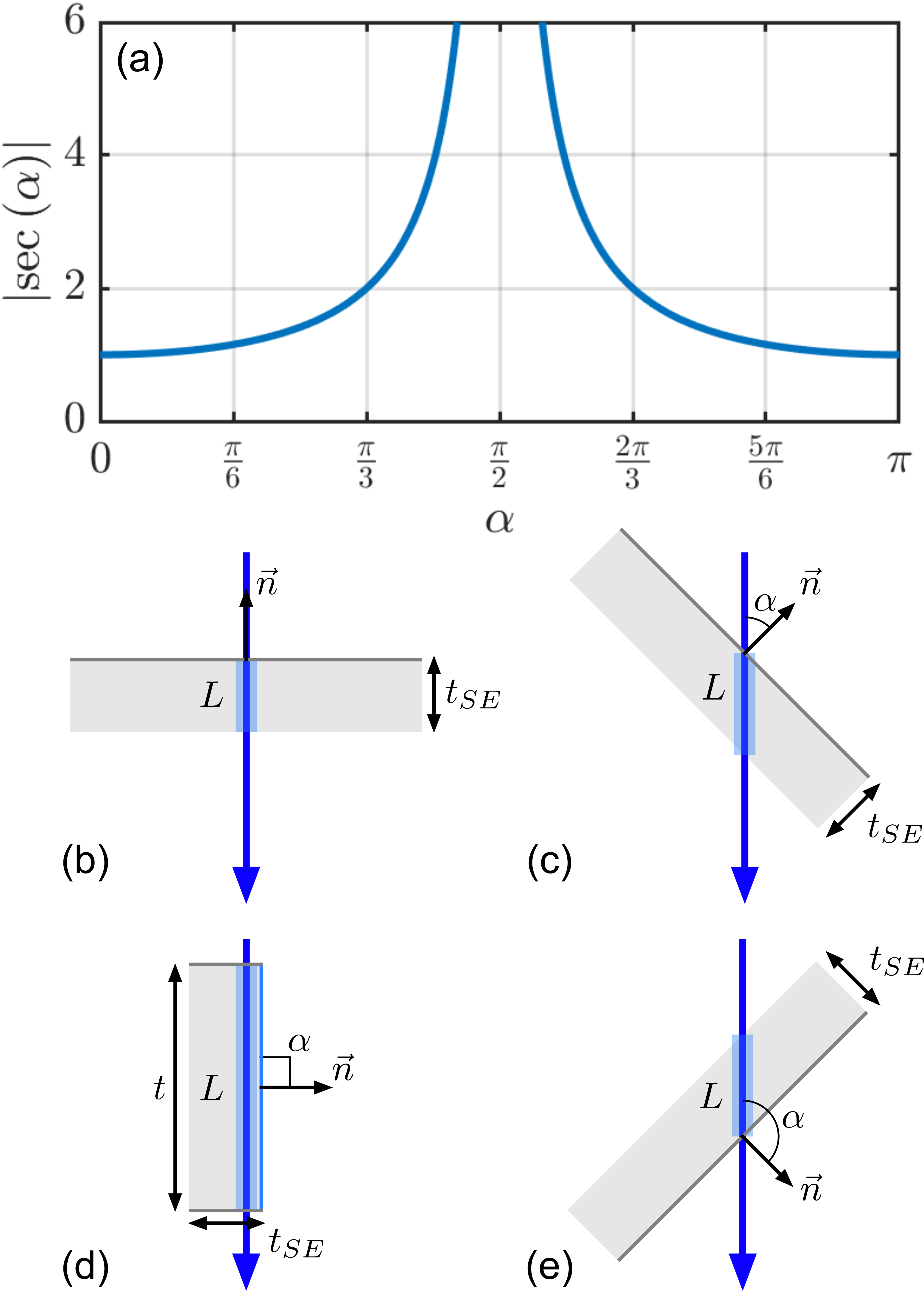}
	\caption{(a) Dependence of SE yield $\delta$ on the surface inclination angle $\alpha$ relative to electron beam incidence in STEM expressed as $\left| \sec{(\alpha)} \right|$. (b-d) Schematic showing the length of the primary electron beam $L$ (highlighted in blue color) within SE escape depth $t_{SE}$ at different surface inclination angles $\alpha$: (b) $\alpha = 0$, (c) $0 < \alpha < \frac{\pi}{2}$, (d) $\alpha = \frac{\pi}{2}$, (e) $\frac{\pi}{2} < \alpha < \pi$.}
	\label{fig:1}
\end{figure}

Similar to SEM, the overall SE intensity measured for each scanning position is determined by the surface inclination angle $\alpha$ relative to beam incidence and angular distribution of emitted SEs~(\textbf{Figure~\ref{fig:1}}). The presence of the entrance and exit surfaces for the primary electron beam, however, extends the $\alpha$ range to higher angles: $\alpha \in \left[0, \pi\right]$. Hereby, it should be noted that $\sec{(\alpha)}$ equals infinity at $\alpha = \frac{\pi}{2}$ making this expression unphysical. Therefore, the problem can be resolved by taking the local thickness of the specimen (thickness of the sample in incidence point of the primary beam) $t$  into account. Namely, the primary electron beam path inside the material can never be longer than $t$ (\textbf{Figure~\ref{fig:1}~d}), such that we find that for each scanning position, the SE yield is defined as:

\begin{equation} \label{eq:Equation 7}
    \delta \propto \min {\left(\frac{t}{t_{SE}}, \left| \sec{(\alpha)} \right|\right)}. 
\end{equation}

Opposed to \textbf{Equation~\ref{eq:Equation 4}}, the dependence of the SE yield $\delta$ on $\alpha$ and $\theta$, then, will transform into:

\begin{equation} \label{eq:Equation 8}
    \frac{d \delta}{d \Omega} = \frac{\delta_{0}}{\pi} \min {\left(\frac{t}{t_{SE}}, \left| \sec{(\alpha)} \right|\right)} \cos{(\theta)}.
\end{equation}

\par

\begin{figure}
	\centering
	\includegraphics[width=1\columnwidth]{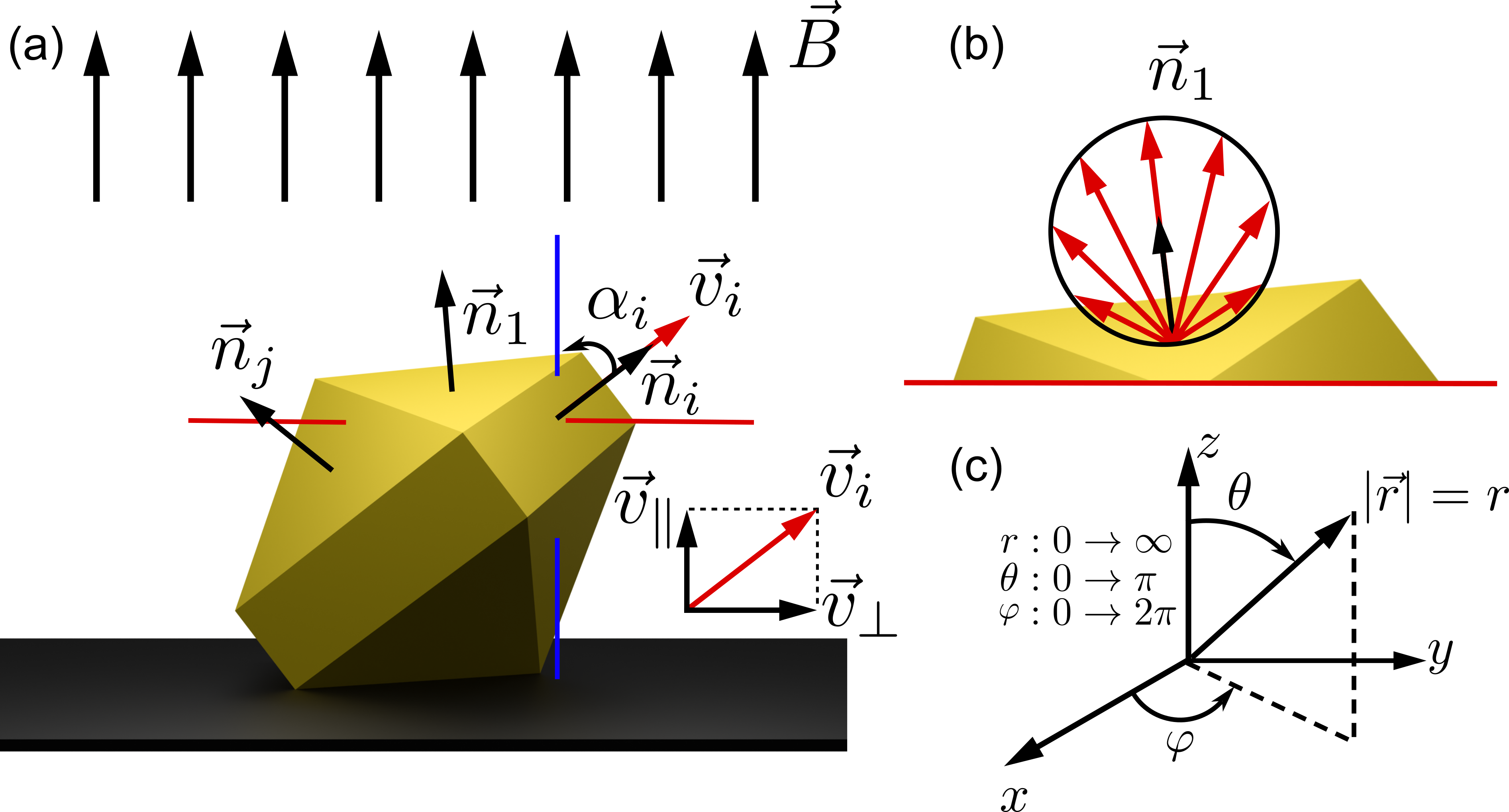}
	\caption{(a) Schematic representation of the imposed problem. (b) Emission of SEs according to Lambert’s law. The lengths of the red arrows in panel (b) are proportional to the number of SEs emitted in their direction, black arrows indicate the surface normals $\vec{n}$. (c) The spherical coordinate system used in the problem solution.}
	\label{fig:2}
\end{figure}

SEs emitted into the vacuum from the NP surface, defined by normal vector $\vec{n}_i$,  in an arbitrary direction determined by velocity vector $\vec{v}_i$ will be affected by the presence of a magnetic field of the objective lens $\vec{B}$~($B \approx$ 2T). This magnetic field is assumed to be homogeneous~(magnetic lines are parallel to each other) since the sample is small compared to the pole piece gap volume. The velocity of the SEs, which is oblique to the magnetic field, can be decomposed into parallel and perpendicular components to the magnetic field $\vec{v}_{\parallel}$ and $\vec{v}_{\perp}$ respectively. $\Vec{v}_{\parallel}$ defines the motion of the SEs along the magnetic field axis in the direction defined by $\vec{v}_{\parallel}$. At the same time, the $\vec{v}_{\perp}$ vector results in a circular motion of the SEs induced by the Lorentz force. Therefore, electrons are spiralling up or down in the magnetic field. SEs with $\vec{v}_{\parallel}$ parallel to $\vec{B}$ are spiralling upwards and reaching the detector in case of off-sample SE-STEM, or contribute to the SEEBIC signal. Conversely, SEs with $\vec{v}_{\parallel}$ antiparallel to $\vec{B}$ are spiralling downwards and being recaptured by the NP itself and support film if present, and thus they do not contribute to the overall SE-STEM (SEEBIC) signal. The fraction of the escaped SEs without being recaptured can be estimated by integrating $\frac{d \delta}{d \Omega}$ over a solid angle $d \Omega$, taking into account boundary conditions governed by the relative orientation of the surface of the NP and the support to magnetic field lines and position of the detector (in case of off-sample SE detection). Considering the symmetry of the problem~(\textbf{Figure~\ref{fig:2}}), the integration can be done in spherical coordinates, associated with the surface normal for every scanning position~(see Supporting Information).

\begin{equation} \label{eq:Equation 9}
    \delta = \frac{\delta_{0}}{\pi} \min{\left(\frac{t}{t_{SE}}, \left| \sec{(\alpha)} \right|\right)} \cdot I, 
\end{equation}

\vspace{5pt}

\noindent{}where $I=\int_{0}^{\varphi \left( \alpha \right)}\,d\varphi \int_{0}^{\theta_m(\varphi(\alpha))}\,d\theta \sin{(\theta)} \cos{(\theta)}$. The upper integration limit over $\varphi$, here, is dependent on inclination angle $\alpha$ and the upper integration limit $\theta_m$ is determined by the intersection of the unit sphere with a plane perpendicular to magnetic field axis $\theta_m = \theta_m(\varphi(\alpha))$~(see Supporting information).

\vspace{5pt}

The solution of the integral $I$ (see Supporting Information) results in a cosine-like law:

\begin{equation} \label{eq:Equation 10}
    I = \frac{\pi}{2}(1+\cos{(\alpha)}), \alpha \in \left[0, \pi\right]
\end{equation}

\begin{figure}
	\centering
	\includegraphics[width=1\columnwidth]{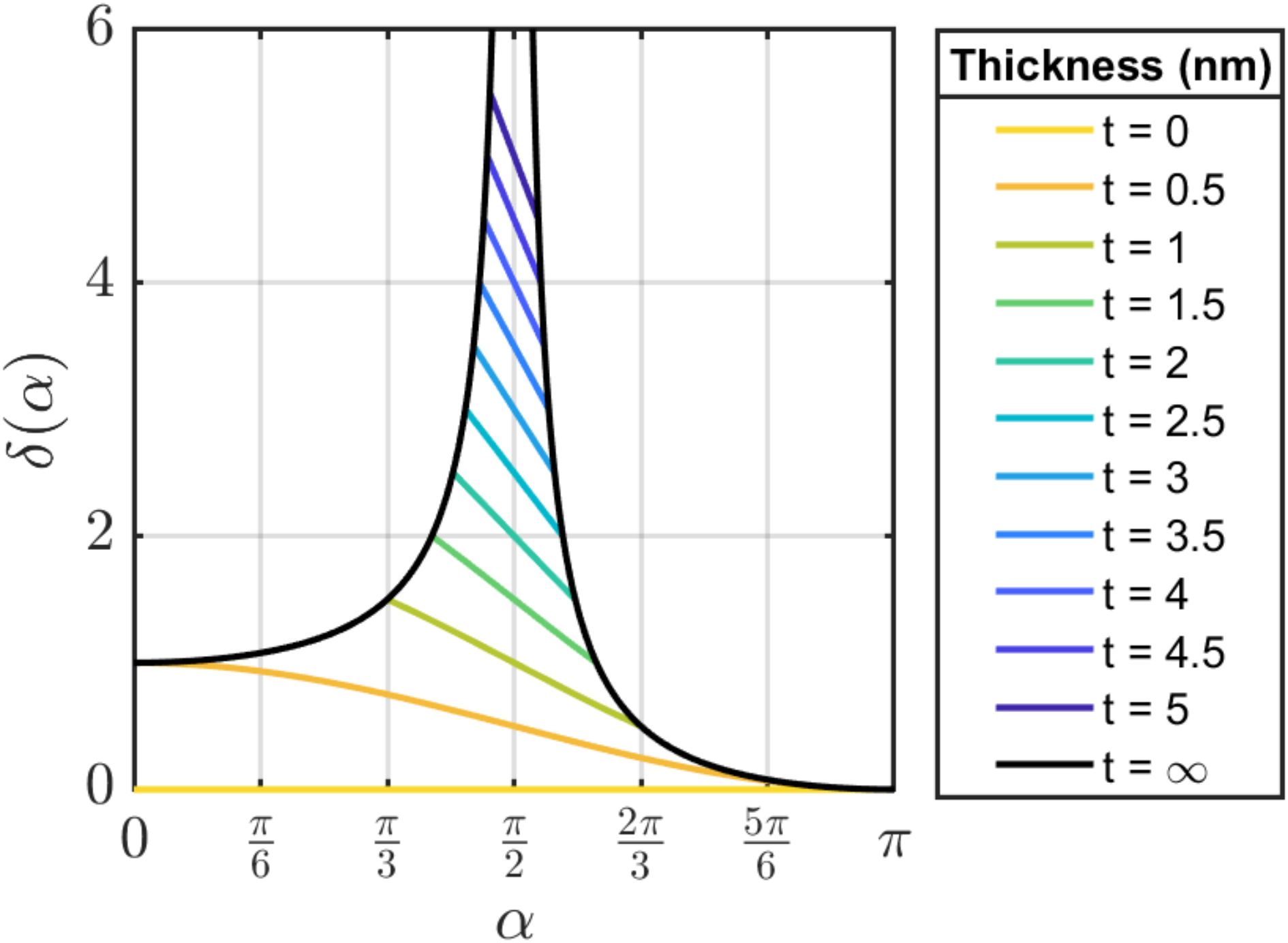}
	\caption{Dependence of SE yield $\delta$ on surface inclination angle $\alpha$ and local thickness $t$ in STEM for gold ($t_{SE}$ = 0.5~nm~\cite{lin2005}) according to \textbf{Equation~\ref{eq:Equation 11}}.}
	\label{fig:3}
\end{figure}

In general, we find that the SE-yield $\delta$ for each surface element is a function of the surface inclination angle $\alpha$ and the local thickness $t$ of the specimen at the scan position (\textbf{Figure~\ref{fig:3}}):

\begin{equation} \label{eq:Equation 11}
    \delta = \frac{\delta_0}{2} \min {\left(\frac{t}{t_{SE}}, \left| \sec{(\alpha)} \right|\right)} (1+\cos{(\alpha)}), \alpha \in \left[0, \pi\right] 
\end{equation}

\section{Results and Discussion}

\begin{figure}
	\centering
	\includegraphics[width=1\columnwidth]{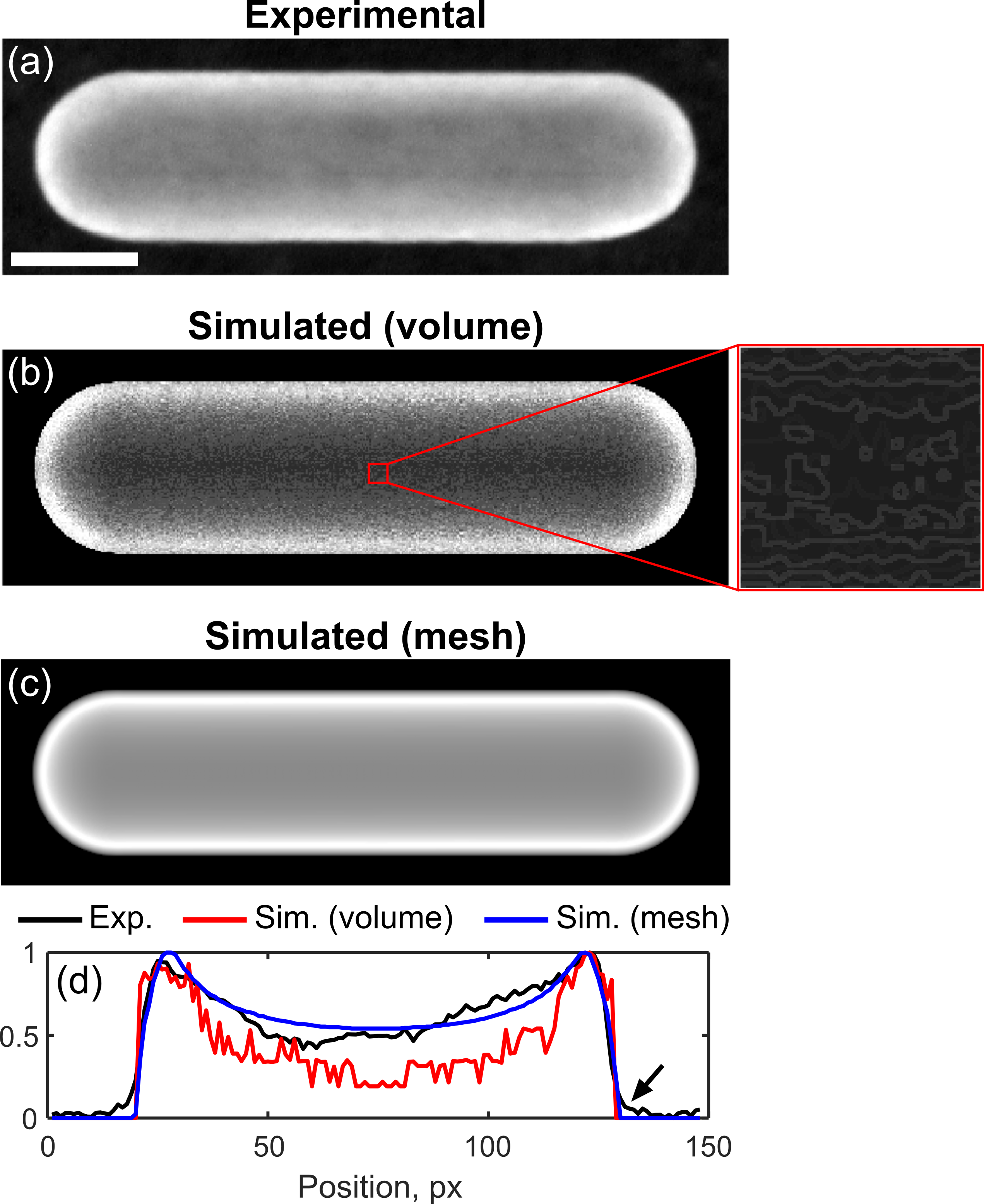}
    \caption{(a) Experimental SEEBIC and (b-c) simulated SE-STEM images of Au nanorod using (b) voxelized and (c) mesh representation of 3D object. (d) Line profile taken across the long axis of the nanorod for experimental and simulated SEEBIC images. Black arrow in panel (d) shows a sharp transition of the contrast between the nanorod and the background. Scale bar is 20~nm.}
	\label{fig:4}
\end{figure}

The proposed physical model for topographical contrast formation can be directly applied to simulate SE-STEM images. We used synthetic 3D models of the nanorod to simulate the SE-STEM image based on voxelized representation of 3D data~(\textbf{Figure~\ref{fig:4}~b}). When working with voxelized volumes, the first step in the simulation algorithm involves converting the data into a binary volume using Otsu thresholding~\cite{otsu1979}. Surface normals are then computed from the 3D image gradient of this binary volume, and the angle $\alpha$ is determined for each voxel at the surface. A thickness map is generated by projecting the volume onto the XY plane, and the SE-STEM image intensity is computed using \textbf{Equation~\ref{eq:Equation 11}}. From the image, it can be appreciated that the simulated SE image qualitatively resembles the experimental one~(\textbf{Figure~\ref{fig:4}~a}) showing bright contrast at the edges of the nanorod. However, a closer look at the simulated data reveals a contour-like pattern in the centre of the nanorod (see inset in panel (b) of \textbf{Figure~\ref{fig:4}}). This pattern is a signature of the staircase artifact arising due to the voxel-based nature of used 3D volumetric data. This effect can be minimized using a finer resolution of the grid (bigger size of the volume) at the cost of increased computational resources or advanced smoothing algorithms~\cite{lempitsky2010} can be applied to the volume.\par 

Since evaluation of the SE yield for voxelized 3D object is computationally demanding, we propose to work with triangular mesh representation of 3D object. The calculation of SE yield can, therefore, be done for each surface element of the NP. Surface normals, can be directly obtained for each mesh element, and the thickness map can also be retrieved. In this case, simulated SE-STEM image of the nanorod qualitatively resembles the experimental SEEBIC image with no presence of the staircase artifacts~(\textbf{Figure~\ref{fig:4}~c}).
\par

Line profiles taken along the long axis of the nanorod demonstrate quantitative correlation. However, the sharp transition of the contrast can be noticed in simulated SE-STEM images (as shown with black arrow in \textbf{Figure~\ref{fig:4}~d}). This can be attributed to the fact that our model does not take into account SEs emitted from the support film, whereas in reality, SEs emitted from the support can escape from under the nanorod contributing to the overall SE intensity, making a contrast transition at the edges of the nanorods smoother. Additionally, this difference can be explained by defocus and finite probe size.

\begin{figure}
	\centering
	\includegraphics[width=1\columnwidth]{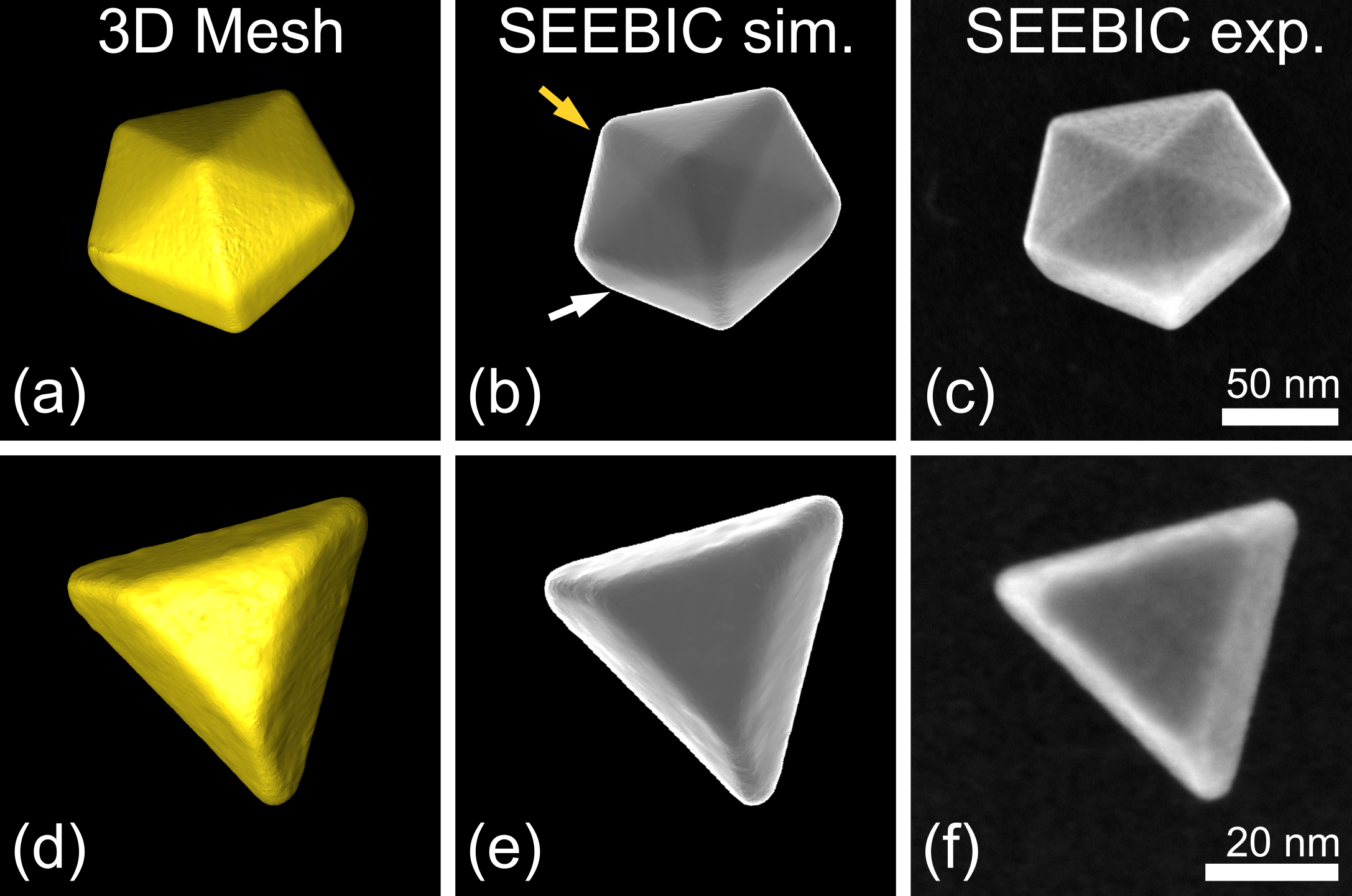}
	\caption{(a, d) 3D triangular meshes obtained from electron tomography reconstruction and corresponding (b, e) simulated SE-STEM and (c, f) experimental SEEBIC images of (a-c) gold Ino decahedron~\cite{ino1969} and (d-f) gold nanotriangle~\cite{scarabelli2014}.}
	\label{fig:5}
\end{figure}

We applied the proposed approach to simulate SE-STEM images from the 3D data obtained from the electron tomography experiment. We used electron tomography reconstructions of a gold Ino decahedron NP and gold nanotriangle. \textbf{Figure~\ref{fig:5}} shows experimental SEEBIC images and simulated data based on electron tomography reconstruction. It can be seen that our model successfully replicates the line of bright contrast at the tip of the decahedron (as shown with a white arrow). Moreover, the simulated image shows areas of dark contrast that correlate with the experimental image (shown with a yellow arrow). The proposed physical model and simulation results can be used as an aid for the interpretation of the contrasts arising in SE images that are not always straightforward to interpret. However, for reliable use of the simulated data, the simulation approach needs to be further improved. Firstly, to account for the re-absorption of SEs by structural features of the NP itself, we need to incorporate a ray tracing solver which will allow us to simulate complex-shaped nanoparticles. Secondly, here, we assume that the magnetic field around the sample is homogeneous, whereas in reality, presense of the lens magnetic line bending leading to recapture of SEs due to so-called magnetic mirror effect. For more accurate model, this effect needs also be taken into consideration since it influences trajectories of emitted SEs (see Supporting information). \par

\begin{figure}[]
	\centering
	\includegraphics[width=1\columnwidth]{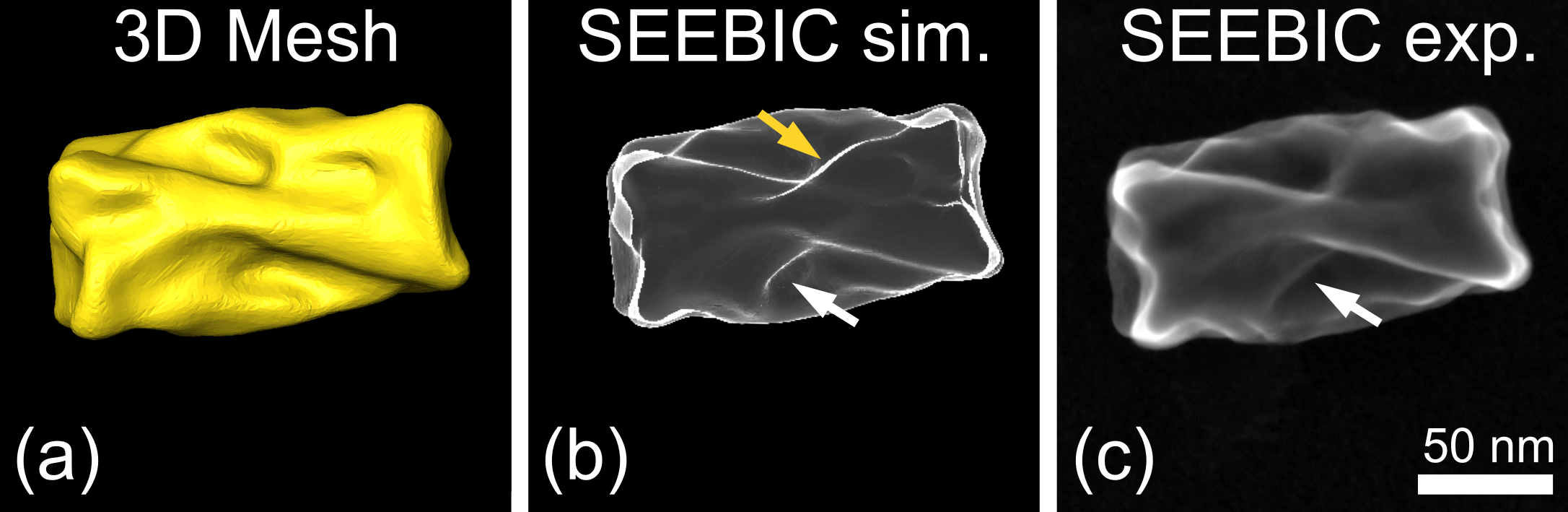}
	\caption{(a) 3D triangular mesh obtained from electron tomography reconstruction and corresponding (b) simulated SE-STEM and (c) experimental SEEBIC images of twisted gold nanorod~\cite{ni2022}.}
	\label{fig:6}
\end{figure}

To demonstrate the limitations of the current approach, we used 3D model of twisted Au nanorod obtained from electron tomography experiment~(\textbf{Figure~\ref{fig:6}}). Since such complex morphology has cavities and ridges formed by the helical twist, our approach is not well suitable since it cannot account for re-absorption of the emitted SEs by the structural features of the NP itself. Simillar to experimental SEEBIC mage, simulated image shows the contrast from the part of the helical ridge at the bottom surface of the NP (see white arrows in \textbf{Figure~\ref{fig:6}}). However, this helical ridge is significantly less pronounced in the experimental images (as indicated by yellow arrow). This can be attribute to the fact that SEs emitted from the sides of the ridge in that area are absorbed by the surface of the nanoparticle. To account for this effect, ray tracing implementation is required. 

It should be noted that in the simulations presented here, we do not distinguish between conventional off-sample SE-STEM and SEEBIC imaging, although the differences in their detection principles may influence the practical aspects of use and contrast formation in different ways.
\par
Off-sample SE-STEM imaging (\textit{e.g.}, using an Everhart-Thornley or a microchannel plate detector) is heavily constrained by geometrical considerations, specifically the location of the detector relative to the sample. Assuming there is no recapture of the SEs by the surface or support film, which is the case for free-standing NP, the detection efficiency ($D$) of SEEBIC can be considered unity, whereas an off-sample detector can only collect SEs that reach it after spiralling in the magnetic field of the objective lens. Typically, an off-sample detector is placed either above or below the sample, although a double detector configuration has been reported~\cite{inada2014}. To account for these geometric effects in simulations, the integral $I$ in \textbf{Equation~\ref{eq:Equation 9}} needs to be re-evaluated taking new boundary conditions into account.
\par
Additionally, unlike off-sample SE-STEM, SEEBIC is inherently insensitive to backscattered electrons and remote SE signals such as SEs generated by backscattered electrons from components inside electron microscope (SE-III and SE-IV~\cite{inada2014}). Although the effect of backscattered electrons is negligible under STEM conditions, at lower acceleration voltages, it can play a significant role and lead to unwanted background signal. To account for this effect Monte-Carlo simulation coupled with an electromagnetic solver is required.
\par
SEEBIC detection is based on the use of a transimpedance amplifier circuit of which the performance, including both signal-to-noise ratio and acquisition speed is fundamentally limited by the gain-bandwidth trade-off (bandwidth is inversely proportional to transimpedance gain). Consequently, the realistic bandwidth of the SEEBIC is typically limited to $\approx$10 kHz. In contrast, in the case of off-sample detection, the signal-to-noise ratio is very high (depending on the number of dynodes in the photomultiplier or in the microchannel plate) with a bandwidth in the MHz range. Although acquisition speed has no direct impact on the image SE image formation, the signal-to-noise ratio might be important for the realistic noise simulation in the images.
\par
Finally, since SEEBIC relies on charge balance within the sample~\cite{hubbard2018}, primary beam absorption must be negligible in comparison to the current associated with SE emission, posing limitations on the sample thickness. On the other hand, off-sample SE-STEM detection is independent of the sample thickness, and even atomic resolution SE-STEM imaging of an 18-$\mu$m-thick sample has been demonstrated in practice recently~\cite{hwang2024}. The effect of the primary beam absorption can be assessed through the simulation of the electron scattering in the material using a Monte-Carlo simulation.

\section{Conclusions}
In this paper, we proposed a forward analytical model for SE topographical contrast formation in STEM, based on the physics of SE emission and the interaction of emitted SEs with the magnetic field of the objective lens. We demonstrated that SE-STEM topographical contrast is a function of both the surface inclination angle and the local thickness of the sample. The physical model was used to simulate SE images using both synthetic 3D data and 3D data obtained from electron tomography experiments, and close agreement with experimental SEEBIC images was found. Finally, we discussed the current limitations of the proposed approach. Future work will focus on further improvement of the model and implementation of the ray tracing solver.

\printcredits

\section*{Declaration of competing interest}
The authors declare that they have no known competing financial interests or personal relationships that could have appeared to influence the work reported in this paper.

\section*{Acknowledgments}

This work was supported by the European Research Council (ERC SyG No. 101166855 CHIRAL-PRO to S.B.). The authors acknowledge financial support from the Research Foundation-Flanders (FWO) through project funding (G0A7723N). The authors thank Prof. Luis M. Liz-Marzán for providing the gold nanoparticle samples used in this work.

\appendix
\section{Supplementary data}
Supplementary material related to this article can be found online at

\bibliographystyle{model1-num-names}

\bibliography{cas-refs}


\end{document}